\documentclass[preprint,showpacs,preprintnumbers,amsmath,amssymb]{revtex4}
\usepackage{amsmath,amssymb,graphics,epsfig,subfigure}

\begin{document}

\title{A note on Friedmann equation of FRW universe in deformed
 Ho\v{r}ava-Lifshitz gravity from entropic force}

\author{Shao-Wen Wei\footnote{E-mail: weishaow06@lzu.edu.cn},
        Yu-Xiao Liu\footnote{E-mail: liuyx@lzu.edu.cn, corresponding author},
        Yong-Qiang Wang\footnote{E-mail: yqwang@lzu.edu.cn}
        }
\affiliation{
    Institute of Theoretical Physics, Lanzhou University,
           Lanzhou 730000, P. R. China}

\begin{abstract}

With entropic interpretation of gravity proposed by Verlinde, we obtain the Friedmann equation of the Friedmann-Robertson-Walker universe for the deformed Ho\v{r}ava-Lifshitz gravity. It is shown that, when the parameter of Ho\v{r}ava-Lifshitz gravity $\omega\rightarrow \infty$, the modified Friedmann equation will go back to the one in Einstein gravity. This results may imply that the entropic interpretation of gravity is effective for the deformed Ho\v{r}ava-Lifshitz gravity.
\end{abstract}

\pacs{ 04.70.Dy \\
   Keywords: Entropic force, Horava-Lifshitz gravity, Friedmann equation
 }

\maketitle

Recently, motivated by Lifshitz theory in solid state physics
\cite{Lifshitz1941}, Ho\v{r}ava proposed a new gravity theory at a
Lifshitz point \cite{Horava2009prd,Horava2009jhep,Horava2009}. The
theory is usually referred to as the Ho\v{r}ava-Lifshitz (HL) theory.
It has manifest 3-dimensional spatial general covariance and time
reparametrization invariance. This is a non-relativistic
renormalizable theory of gravity and recovers the four dimensional
general covariance only in an infrared limit. Thus, it may be
regarded as a UV complete candidate for general relativity.

The black hole solutions in this gravity theory have been attracted
much attention. L$\ddot{u}$-Mei-Pope (LMP) first obtained the
spherically symmetric black hole solution with dynamical parameter
$\lambda$ in asymptotically Lifshitz spacetimes \cite{Lu2009prl}.
Following, other black hole solutions and cosmological solutions were
obtained
\cite{cai2009prd,cai2009jhep,Colgain2009jhep,Kehagias2009plb,park2009jhep,
Ghodsi2009,Lee2009,Tang2009,Setare2009,Kiritsis2009,Tang20092,
Beato2010,Polychronakos2009,Kiritsisb2009}. The thermodynamic
properties and dynamical properties of different black hole solutions
were also investigated intensively
\cite{Cao2009plb}-
%,bMyung2009plb,bChen2009plb,Mann2009jhep,Myunge2009plb,Myungh2009plb,Chen2009,
%Peng2009,aMyung2009,Ohta2009,Chen2010,wang2010,bMyung2009,Majhi2009,
%Wangd2009,Varghese2009,Wu2009,Myungf2009,Wang2009,Castillob2009,Setare2010,Gao2010,Gao20102,
%Gao20103,Duttajcap2010
\cite{Saridakis2010cqg}.

On the other hand, Verlinde \cite{Verlinde} recently made a robust
suggestion that gravity can be explained as an entropic force caused
by the changes in the information associated with the positions of
material bodies. This idea implies that gravity is not fundamental.
In fact, the similar idea can be traced back to the work given by
Sakharov about forty years ago \cite{Sakharov}. In the original
article, with the assumption of the entropic force and the Unruh
temperature, Verlinde obtained the second law of Newton mechanics.
With the holographic principle and the equipartition law of energy,
Verlinde also derived the Newtonian law of gravitation. On the other
side, using the equipartition law of energy for the horizon degrees
of freedom together with the thermodynamic relation $S=E/(2T)$,
Padmanabhan also obtained the Newton's law of gravity
\cite{Padmanabhan1}.

With the idea of entropic force, some work has been carried out
recently. At almost the same time, Shu and Gong \cite{Shu}, Cai, Cao
and Ohta \cite{caid} obtained the Friedmann equations, respectively.
Smomlin derived the Newtonian gravity in loop quantum gravity
\cite{Smolin}. On the other hand, Makela pointed out that Verlinde's
entropic force is actually the consequence of a specific microscopic
model of spacetime \cite{Makela}. Caravelli and Modesto also applied
the similar ideas to the construction of holographic actions from
black hole entropy \cite{Caravellila}. In \cite{Li}, Li and Wang
showed that the holographic dark energy can be derived from the
entropic force formula. Gao \cite{Gao} gave a modified entropic force
in the Debye's model. Wang \cite{Wang} regarded the coulomb force as
an entropic force. Zhang, Gong and Zhu \cite{Zhang} discussed three
different corrections to the area law of entropy and obtained the
modified Friedmann equations. Other applications can be seen in
\cite{Piazza,Culetu,Wang2}.

In this paper, we will study the Friedmann equation of the Friedmann-Robertson-Walker (FRW) in deformed HL gravity from the view of entropic force. Let us start with an (3 +1)-dimensional 	 
FRW universe, which is described by the metric
\begin{equation}
 ds^{2}=-dt^{2}+a^{2}(t)\,\left(\frac{dr^{2}}{1-kr^{2}}
  +r^{2}d\Omega_{2}^{2}\right),\label{frwmetric}
\end{equation}
where $k$ denotes the spatial curvature constant with $k=+1,\,0$ and
$-1$ corresponding to a closed, flat and open universe, respectively.
$d\Omega_{2}^{2}$ is the line element of a two-dimensional unit
sphere. Adopting a new coordinate $\tilde{r}=a(t)r$, the metric
(\ref{frwmetric}) will become
\begin{equation}
 ds^{2}=h_{ab}dx^{a}dx^{b}+\tilde{r}^{2}d\Omega_{2}^{2}
\end{equation}
with $x^{0}=t$, $x^{1}=r$ , $h_{ab}=$diag$(-1,\, a^{2}/(1-kr^{2}))$.
The dynamical apparent horizon $\tilde{r}_{A}$ is determined by
$h^{ab}\partial_{a}\tilde{r}\partial_{b}\tilde{r}=0$, and is given by
\begin{equation}
\tilde{r}_{A}=\frac{c}{\sqrt{H^{2}+k/a^{2}}},\label{radius2}
\end{equation}
where $H\equiv\dot{a}/a$ is the Hubble parameter and $c$ is the speed of light. The overdot stands for the derivative with respect to the cosmic time $t$. Suppose
that the energy-momentum tensor $T_{\mu\nu}$ of the matter in the
universe has the form of a perfect fluid
$T_{\mu\nu}=(\rho+p)U_{\mu}U_{\nu}+pg_{\mu\nu}$, where $U^{\mu}$
denotes the four-velocity of the fluid and $\rho$ and $p$ are the
energy density and pressure, respectively. The energy conservation
law $\nabla_{\mu}T^{\mu\nu}=0$ gives the continuity equation in the
form
\begin{equation}
 \dot{\rho}+3H\,(\rho+p)=0.\label{continuityequation}
\end{equation}
The time-component of the Einstein equation is nothing but the
standard Friedmann equation
\begin{equation}
 H^{2}+\frac{k}{a^{2}}=\frac{8\pi G}{3}\rho,\label{FriedmannEq}
\end{equation}
which describes the dynamical evolution of the universe model. It is
proved in \cite{Caikim200jhep} that the Friedmann equation
(\ref{FriedmannEq}) can be derived by the first law of
thermodynamics. Much work in this field has been made
\cite{Akbar2006plb,Caicae2008jhep,cai2009cqg,cai2007ptps,Gong2007prl,Zhu2009plb}.
Assuming that the entropy/area relation in Ho\v{r}ava-Lifshitz gravity is $S=A/4$,
the standard Friedmann equation was obtained in \cite{Chen2010} from the
clausius relation.

Next, we will devote our efforts to obtain the modified Friedmann equation
in the deformed HL gravity from the entropic force. First, we assume that the temperature corresponding to the apparent horizon is
\begin{equation}
 T=\frac{\hbar c}{2\pi k_{B}\tilde{r}_{A}}.
\end{equation}
The entropy $S$ depends on the gravity theory and takes different
forms for different gravity theories. In the deformed HL gravity, it
has the form
\begin{equation}
 S=\frac{k_{B}Ac^{3}}{4G\hbar}+\frac{k_{B}\pi}{\omega} \ln \frac{Ac^{3}}{G\hbar}.\label{entropyarea2}
\end{equation}
This entropy/area relation has a logarithmic term, which is a characteristic of HL gravity theory. However, as the parameter $\omega\rightarrow \infty$, it will go back to the one in Einstein gravity. From the definition of the apparent horizon (\ref{radius2}), a straightforward calculation gives
\begin{equation}
 \dot{\tilde{r}}_{A}=-\frac{1}{c^{2}}H\tilde{r}_{A}^{3}
           \left(\dot{H}-\frac{k}{a^{2}}\right).\label{tradius2}
\end{equation}
Following \cite{Hayward1998cqg}, the work density $W$ and
energy-supply vector $\Phi_a$ for our case can be calculated as
follows:
\begin{eqnarray}
 W&=&-\frac{1}{2}T^{ab}h_{ab}=-\frac{1}{2}(p-\rho),\\
 \Phi_{a}&=&T^{b}_{a}\partial_{b}\tilde{r}=-\frac{1}{2}(p+\rho)H\tilde{r}dt+\frac{1}{2}(p+\rho)adr,
\end{eqnarray}
with $T_{ab}$ the projection of the (3 + 1)-dimensional
energy-momentum tensor $T_{\mu\nu}$ of matter in the normal direction
of 2-sphere. Then the amount of energy crossing the apparent horizon
during the time internal $dt$ is \cite{Caikim200jhep}
\begin{eqnarray}
 d E=A\Phi=A(\rho+p)H\tilde{r}_{A}dt.\label{energychange}
\end{eqnarray}

Now, let us turn to the entropic force. Verlinde \cite{Verlinde} made an interesting suggestion that gravity may be regarded as an entropic force caused by the
changes in the information associated with the positions of material
bodies. With this powerful assumption together with the holographic
principle and the equipartition law of energy, Verlinde obtained the
second law of Newtonian mechanics and Newtonian law of gravitation.
The essence of the idea is that gravity is not fundamental but is an
entropic force, which can be expressed in the following form
\begin{eqnarray}
 F\Delta x=T\Delta S,
\end{eqnarray}
where $\Delta S$ denotes the small change of entropy $S$ of the
gravitational system, $\Delta x$ is the displacement of a particle in
the gravitational system, $T$ is the temperature of the system, and
$F$ is named entropic force. In the original paper, the author
suggested that the maximal storage space for information on the holographic screen, or total number of bits, is proportional to the area $A$ with the holographic
principle is held, i.e.,
\begin{eqnarray}
 N=\frac{A c^{3}}{G \hbar},\label{number}
\end{eqnarray}
where $A$ is area of the holographic screen. Then, with the
equipartition law of energy $E=\frac{1}{2}Nk_{B}T$ and $E=Mc^{2}$
($M$ represents the mass that would emerge in the part of space
enclosed by the holographic screen), one will recover the Newton's
law of gravitation:
\begin{eqnarray}
 F=G\frac{Mm}{R^{2}},
\end{eqnarray}
where the relations $F=ma$ ($a$ is regarded as the acceleration
corresponding to the Unruh temperature) and $A=4\pi R^{2}$ are used.

Here, we propose that the total number of bits for the information
is proportional to the entropy $S$ rather than the area $A$ in a gravity
theory as did in \cite{Zhang}. Thus, for a general gravity theory, one could rewrite
(\ref{number}) as \footnote{In a general gravity theory, entropy $S$
is proportional to the total number of microstates $\Omega$. For
$N\propto \ln \Omega$, so it is easy to obtain $N\propto S$. On the
other hand, in the frame of Einstein gravity, we have $S\propto A$.
Then it is straightforward to obtain $N\propto A$. This is also
implied in the work \cite{Padmanabhan}.}
\begin{eqnarray}
 N=\frac{4S}{k_B}. \label{number2}
\end{eqnarray}

Next, we will reproduce the Friedmann equation from this entropic
force method. With (\ref{number2}), we obtain the total number of bits for the HL gravity
\begin{eqnarray}
 N=\frac{Ac^{3}}{G\hbar}+\frac{4\pi}{\omega} \ln \frac{Ac^{3}}{G\hbar}.\label{mumber3}
\end{eqnarray}
From the view of entropic force, the energy $E$ for a thermodynamic system can be reflected by the number of bits on the holographic screen through the equipartition
law $E=\frac{1}{2}Nk_{B}T$ with $T$ the temperature of the holographic screen. Here, we choose the apparent horizon to be the holographic screen. So the temperature $T$ is that for the apparent horizon. Then, we can get the change of the energy $dE$ from the equipartition law
\begin{eqnarray}
 dE&=&\frac{1}{2}Nk_{B}dT+\frac{1}{2}k_{B}TdN %\nonumber\\
   =\frac{8G\hbar\pi+  A \omega  c^3-4G\hbar\pi\ln
             \left(\frac{A c^3}{G \hbar}\right) }
           {4 A^{3/2}G \sqrt{\pi } \omega }c dA.
   \label{energychange2}
\end{eqnarray}
Comparing (\ref{energychange}) and (\ref{energychange2}), we reach
\begin{eqnarray}
 %&&
 \bigg[8G\hbar\pi+  A \omega  c^3-4G\hbar\pi
  \ln\left(\frac{A c^3}{G \hbar}\right)\bigg]d\tilde{r}_{A}
 %  \nonumber  \\  &&
 =16\pi^{2}cG\omega \tilde{r}^{5}_{A} (\rho+p)Hdt.
\end{eqnarray}
With the use of (\ref{tradius2}) and the continuity equation
(\ref{continuityequation}), we get
\begin{eqnarray}
 %&&
 \bigg[8G\hbar\pi+  A \omega  c^3-4G\hbar\pi
  \ln\left(\frac{A c^3}{G \hbar}\right)\bigg]
  \left(H-\frac{k}{a^{2}}\right)H
 %   \nonumber  \\  &&
    =\frac{16G\omega c^{3}\pi^{2}\tilde{r}^{2}_{A}}{3}\dot{\rho}. \label{frwentropic}
\end{eqnarray}
Using the relation
\begin{eqnarray}
 2H\left(\dot{H}-\frac{k}{a^{2}}\right)
 =\frac{d}{dt}\left(H^{2}+\frac{k}{a^{2}}\right)
\end{eqnarray}
 and integrating (\ref{frwentropic}), we derive the following equation
\begin{eqnarray}
 %&&
 \frac{\hbar G}{4\omega c^{5}}
    \left(H^{2}+\frac{k}{a^{2}}\right)^{2}
    \left[3 -2\ln\bigg(\frac{4\pi c^{5}}
           {G\hbar(H^{2}+{k}/{a^{2}})}\bigg)
    \right] %~~~~~~~ \nonumber  \\ &&
 + \left(H^{2}+\frac{k}{a^{2}}\right)
 =\frac{8\pi G}{3}\rho,\label{frwentropic2}
\end{eqnarray}
where the integration constant has also been absorbed into the energy
density $\rho$. Eqs. (\ref{frwentropic}) and (\ref{frwentropic2}) are
the modified Friedmann equations from the entropic force. Comparing with the standard Friedmann equation (\ref{FriedmannEq}), the modified one is in a complicated form. This means that in principle we can distinguish the HL gravity from the Einstein one by exploring its Friedmann equation. As we know, when the parameter $\omega$ of the HL gravity approaches infinite, the Einstein gravity will be recovered. Setting $\omega\rightarrow \infty$, we can see that the first term in (\ref{frwentropic2}) vanishes and the standard Friedmann equation will be obtained, which is exactly consistent with the result in \cite{Chen2010}.

%\section{Sumarry}

In this letter, we obtain the modified Friedmann equation of FRW
universe in the deformed HL gravity theory from the entropic force. In the
deformed HL gravity, the entropy/area relation is generally
considered to have a logarithmic term, i.e.,
$S=\frac{A}{4G}+\frac{\pi}{\omega}\ln A$. The parameter $\omega$ can
be regarded as a characteristic parameter in the deformed HL gravity
and the entropy/area relation will recover to Bekenstein-Hawking
entropy/area law when $\omega\rightarrow \infty$. To some extent,
this logarithmic term represents a feature of black holes in the
deformed HL gravity. Considering the apparent horizon of the universe is a
holographic screen, we propose that the total number of bits for the information of the screen is $N=\frac{4S}{k_B}$. Then the energy included in the screen can be described by the equipartition law of the energy. Thus we obtain the modified Friedmann equation, in which a square term
of $(H^{2}+\frac{k}{a^{2}})$ and a logarithmic term appeared, which provide a way to distinguish the HL gravity from the Einstein one. However, as the parameter $\omega\rightarrow \infty$, the modified equation will become the standard one. Our result may be suggested that gravity indeed has a thermodynamic origin for a non-Einstein gravity and it may be useful for further understand of the
holographic properties and Friedmann equation for other gravity theories.

%\section*{Acknowledgement}

Y.X. Liu would like to thank Dr. Li-Ming Cao for his helpful
discussions. This work was supported by the Program for New Century Excellent
Talents in University, the National Natural Science Foundation of China (No. 11075065), the Fundamental Research Funds for the Central Universities (No. lzujbky-2009-54, No. lzujbky-2009-163 and lzujbky-2009-122),
and the Fundamental Research Fund for Physics and Mathematics of Lanzhou University (No. LZULL200912).

\end{document}